\documentclass[notitlepage,reprint,amsmath,amssymb,aps,prl,onecolumn,superscriptaddress,longbibliography]{revtex4-1}

\usepackage{tikz}
\usepackage{multirow}
\usepackage{graphicx}
\usepackage{dcolumn}
\usepackage{bm}
\usepackage{array}
\usepackage{booktabs}
\usepackage[unicode=true,colorlinks=true,citecolor=blue,urlcolor=blue]{hyperref}

\newcommand{\modtab}[3]
{
	\begin{tabular}[c]{@{}c@{}}
		\specialrule{0em}{1pt}{#1}
		#3\\
		\specialrule{0em}{1pt}{#2}
	\end{tabular}
}

\begin{document}
	
	\preprint{APS/123-QED}
	
	\title{Non-Hermitian skin effect and delocalized edge states in photonic crystals with anomalous parity-time symmetry}
	
	\author{Qinghui Yan}
	\affiliation{
		Interdisciplinary Center for Quantum Information, State Key Laboratory of Modern Optical Instrumentation, ZJU-Hangzhou Global Scientific and Technological Innovation Center, Zhejiang University, Hangzhou 310027, China.
	}
	\affiliation{
		International Joint Innovation Center, Key Lab. of Advanced Micro/Nano Electronic Devices \& Smart Systems of Zhejiang, The Electromagnetics Academy at Zhejiang University, Zhejiang University, Haining 314400, China.
	}
	
	\author{Hongsheng Chen}
	\email{hansomchen@zju.edu.cn}
		\affiliation{
		Interdisciplinary Center for Quantum Information, State Key Laboratory of Modern Optical Instrumentation, ZJU-Hangzhou Global Scientific and Technological Innovation Center, Zhejiang University, Hangzhou 310027, China.
	}
	\affiliation{
		International Joint Innovation Center, Key Lab. of Advanced Micro/Nano Electronic Devices \& Smart Systems of Zhejiang, The Electromagnetics Academy at Zhejiang University, Zhejiang University, Haining 314400, China.
	}
	
	\author{Yihao Yang}
	\email{yangyihao@zju.edu.cn}
		\affiliation{
		Interdisciplinary Center for Quantum Information, State Key Laboratory of Modern Optical Instrumentation, ZJU-Hangzhou Global Scientific and Technological Innovation Center, Zhejiang University, Hangzhou 310027, China.
	}
	\affiliation{
		International Joint Innovation Center, Key Lab. of Advanced Micro/Nano Electronic Devices \& Smart Systems of Zhejiang, The Electromagnetics Academy at Zhejiang University, Zhejiang University, Haining 314400, China.
	}
	
	\begin{abstract}
		Non-Hermitian skin effect denotes the exponential localization of a large number of eigen-states in a non-Hermitian lattice under open boundary conditions. Such a non-Hermiticity-induced skin effect can offset the penetration depth of in-gap edge states, leading to counterintuitive delocalized edge modes, which have not been studied in a realistic photonic system such as photonic crystals. Here, we analytically reveal the non-Hermitian skin effect and the delocalized edge states in Maxwell’s equations for non-Hermitian chiral photonic crystals with anomalous parity-time symmetry. Remarkably, we rigorously prove that the penetration depth of the edge states is inversely proportional to the frequency and the real part of the chirality. Our findings pave a way towards exploring novel non-Hermitian phenomena and applications in continuous Maxwell’s equations.
	\end{abstract}

	\maketitle
	
	\section{Introduction}
	Non-Hermitian skin effect (NHSE), the exponential localization of an extensive number of eigen-states at the open boundaries of the system, has gained rapidly growing attention in non-Hermitian physics. Initially proposed in condensed-matter systems under tight-binding approximation, the NHSE has been theoretically constructed and experimentally realized across multiple disciplines~\cite{xiao2020non,helbig2020generalized,PhysRevLett.124.066602,PhysRevResearch.2.023173,PhysRevLett.122.237601,Longhi19,deng2021nonhermitian,braghini2021non,Gao2020,PhysRevApplied.14.064076,wang2021generating}. So far, most studies have been focused on the NHSE in one-dimensional (1D) systems, where the conventional bulk-edge correspondence principle is violated. To recover the bulk-edge correspondence, the generalized Brillouin zone (GBZ) has been proposed, in which a nonzero imaginary part is assigned to the Bloch wavevector $k$~\cite{PhysRevLett.121.086803}. The relation between $Re(k)$ and $Im(k)$ is thus the GBZ, with $Re(\cdot)$ and $Im(\cdot)$ denoting the real and imaginary part of the inputs, respectively. Moreover, GBZ provides an intuitive picture to understand the penetration depths of the localized states. 
	
	The NHSE has challenged our conventional understanding of bulk/edge states, as the bulk states can be localized at the edge, whereas the edge states can be fully delocalized into states that extended in the bulk~\cite{PhysRevB.103.195414}. However, they can still be distinguished by the number of eigen-states; for a lattice sized $N$, the number of bulk states is $O(N)$, while the number of edge states is $O(1)$~\cite{PhysRevLett.124.086801,PhysRevLett.125.126402}. In the context of photonics, the delocalized edge states (DESs) have potentials in applications. For example, due to the frequency isolation and energy delocalization, the DESs manifest themselves as ideal resonant modes for single-mode lasing. While most previous studies on NHSE and DESs have been restricted to tight-binding models, realistic photonic systems such as photonic crystals usually cannot be modeled discretely. It thus requires a continuum approach based on the Maxwell’s equations, which have not yet been studied. 
		
	Here, we analytically study the NHSE and the DESs in 1D non-Hermitian chiral PhCs based on the first-principle calculation by Maxwell’s equations. We start with a Hermitian PhC, and derive the eigen-spectra under periodic boundary condition (PBC) and open boundary condition (OBC), respectively. By adopting the transfer matrix method (TMM), we find that by varying the non-Hermitian chirality of one type of layers, we precisely control the penetration depth of NHSE, resulting in DESs. Finally, we prove the non-Hermitian chiral material satisfies anomalous parity-time symmetry, and discuss how the real and imaginary parts of the chirality affect the dispersion relations and mode profiles, respectively. 

	\begin{figure}
	\centering
	\includegraphics[scale=1]{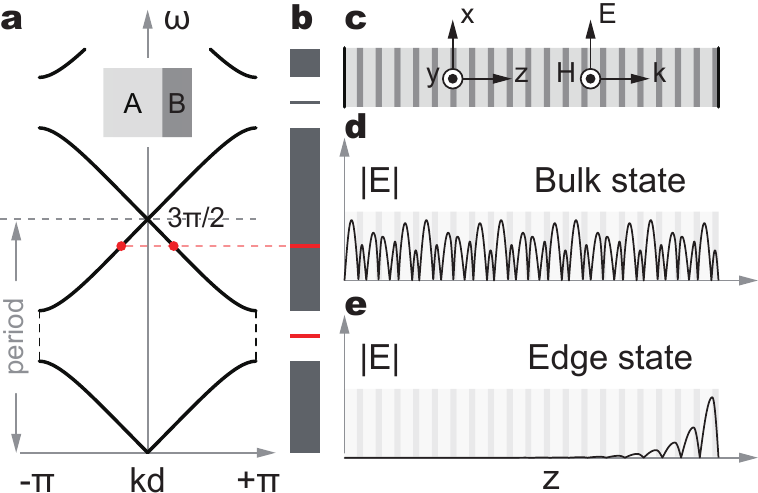}
	\caption{
		Hermitian PhCs. (a) Dispersion relation of PhCs under PBC. By judiciously designing the unit-cell, all bands are periodically arranged in the frequency domain, containing one bandgap in each period. The inset shows the unit-cell. (b) OBC spectra in an FP cavity. In the middle of each bandgap, there is an edge state localized at the right boundary. (c) Schematic of the PhCs under OBC. (d) Mode profile of a bulk state, shown as in-band red line in (a). (e) Mode profile of the in-gap edge state localized at the right boundary. 
	}
	\label{Fig:PhC}
\end{figure}

	\section{Hermitian photonic crystals}
	We start with a piece of Hermitian 1D PhCs with a unit-cell consisting of two layers, namely layer A and layer B. The thickness of two layers are $d_A=2d_B=\frac{2}{3}d$, where $d$ is the lattice constant normalized to 1. The constitutive parameters are $\epsilon_B=4$ and $\epsilon_A=\mu_A=\mu_B=1$, respectively. For simplicity, we suppose $\epsilon_0=\mu_0=c=1$, and only consider the TE polarization with a magnetic field out of plane, as shown in Fig.~\ref{Fig:PhC}(c). Analytically, we obtain the dispersion relation of the PhCs~\cite{Xiao2014}, 
	\begin{equation}
		cos[kd]=\frac{\eta+1}{2}cos[\omega(n_Ad_A+n_Bd_B)]+\frac{-\eta+1}{2}cos[\omega(n_Ad_A-n_Bd_B)]
		\label{Eq:ABSlabDispersion}
	\end{equation}
	The band structure of the PhC under PBC is shown in Fig.~\ref{Fig:PhC}(a). Here, $\eta=(Z_A/Z_B+Z_B/Z_A)/2$, and $Z_{A,B}$ and $n_{A,B}$ are the impedances and refractive indices of layer A and B, respectively. For isotropic materials, $Z=\sqrt{\mu/\epsilon}$ and $n=\sqrt{\epsilon\mu}$, so $n_B=2$, $Z_B=1/2$ and $n_A=Z_A=1$, and thus $\eta=5/4$. Note that $n_Ad_A-n_Bd_B=0$, which simplifies Eq.~\ref{Eq:ABSlabDispersion} to make all bands periodically arranged in the frequency domain, with a normalized periodicity $\omega_0/(c/d)=3\pi/2$. Each period contains two bands and a bandgap in between, as shown in Fig.~\ref{Fig:PhC}(a). 
	
	Next, we study the OBC eigen-spectra. Consider a Fabry-P\'{e}rot~(FP) cavity filled with such PhCs. The cavity is bounded by perfect metals that shield the electric field, analogous to the OBC in lattice models. As shown in Fig. 1(b), the OBC spectra are almost the same as PBC counterparts, except several frequency isolated states in the middle of the bandgap, known as the in-gap edge states. In our example, the finite-size PhC starts with layer A and ends with layer B, so there is only one edge state localized at the right boundary. Because of the OBC, every eigen-state is a standing wave that interferes destructively at the boundaries, mostly by two bulk states under PBC with opposite wavevectors, shown as the solid red dots and the in-band red line in Fig.~\ref{Fig:PhC}(a), and the mode profile in Fig.~\ref{Fig:PhC}(d). Besides most bulk states, an OBC state can destructively interfere by itself, such as the edge state that locates in the bandgap, having no correspondence to the PBC states. Substituting $\omega=\omega_0/2$ into Eq.~\ref{Eq:ABSlabDispersion}, we have $k=\pi+ig_0$, where $g_0=-acosh~\eta$. Here, $g_0<0$ implies the rightward localization of edge state. The penetration depth is $\delta_0=1/|g_0|\approx1.44$ periods. 
	
	\section{Generalized Brillouin zone calculated from transfer matrix method}
	For future analysis, we briefly recall TMM to describe our model. The electromagnetic field in each layer consists of two planewaves propagating forward and backward, described by the TE component of the Maxwell’s equations 
	\begin{equation}
		k\begin{bmatrix} 0 & 1 \\ 1 & 0 \end{bmatrix}
		\begin{bmatrix} E_x \\ H_y \end{bmatrix}
		=\omega
		\begin{bmatrix} \epsilon_{xx} & \xi_{xy} \\ \zeta_{yx} & \mu_{yy} \end{bmatrix}
		\begin{bmatrix} E_x \\ H_y \end{bmatrix} \text{.}
	\end{equation}
	Rewrite it as $n\sigma_1\Psi=\mathcal{M}\Psi$ for brief notations, where $n=k/\omega$ is the definition of refractive index and $\sigma_1$ is the first Pauli matrix. By solving this generalized eigen-value problem, we have two eigen-values $n_\pm$ and two eigen-vectors $\Psi_\pm$, respectively, where the plus-minus sign denotes the forward/backward propagation direction. Arrange $\Psi_+$ and $\Psi_-$ in columns, we have
	\begin{equation}
		\tilde{\Psi}=\begin{bmatrix} \Psi_+ & \Psi_- \end{bmatrix}=\begin{bmatrix} Z_+ & Z_- \\ 1 & -1 \end{bmatrix} \text{.} 
	\end{equation} 
	In this work, we suppose $Z_+=Z_-$ and suppress the plus/minus subscript of impedances. From the definition of $\tilde{\Psi}$ we have $\sigma_1\tilde{\Psi}\tilde{n}=\mathcal{M}\tilde{\Psi}$, where $\tilde{n}=\begin{bmatrix} \begin{smallmatrix} n_+ &  \\  & -n_- \end{smallmatrix} \end{bmatrix})$, which finally leads to 
	\begin{equation}
		\mathcal{M}=\sigma_1\tilde{\Psi}\tilde{n}\tilde{\Psi}^{-1} \text{.} 
		\label{Eq:MaterialMatrix}
	\end{equation}
	In doing so, we can calculate the refractive index and impedance from the constitutive parameters, and vice versa. Take layer A as an example, the electromagnetic field in inside the layer is expressed by 
	\begin{equation}
		\begin{bmatrix}	E_x \\ H_y \end{bmatrix} =
		\begin{bmatrix} Z_A & Z_A \\ 1 & -1 \end{bmatrix}
		\begin{bmatrix}	m_+ \\ m_- \end{bmatrix}\Rightarrow\Psi=\tilde{\Psi}_A\Phi \text{,} 
		\label{Eq:PsiModeConvert}
	\end{equation}
	where $m_+$ and $m_-$ are the coefficients of planewaves. Given the thickness $d_A$ of the layer, we have 
	\begin{equation}
		\Phi_{d_A} =
		\begin{bmatrix} e^{+i\omega n_{A+}d_A} &  \\  & e^{-i\omega n_{A-}d_A} \end{bmatrix}
		\Phi_{0} \Rightarrow \Phi_{d_A}=\textbf{T}_A\Phi_0 \text{,} 
		\label{Eq:PhiPhiConvert}
	\end{equation}
	where the subscript denotes the position of the wavefunction. $n_{A\pm}$ denotes the refractive index for the forward/backward component, which are not the same in chiral materials. Given that the electromagnetic field is continuous along the interface between every two neighboring layers, we have 
	\begin{equation}
		(\tilde{\Psi}_B\textbf{T}_B\tilde{\Psi}_B^{-1})
		(\tilde{\Psi}_A\textbf{T}_A\tilde{\Psi}_A^{-1})
		\Psi_0= \Psi_d = \beta\Psi_0
		\label{Eq:DispersionRelation}
	\end{equation}
	for the unit-cell under Floquet boundary condition, where $\beta=e^{ikd}$ is the generalized phase shift per unit-cell. Here, the \emph{generalized} means $k$ can have nonzero imaginary part to amplify/attenuate the amplitude of the wave. By solving this eigen-value problem, we find two eigen-values, namely $\beta_+$ and $\beta_-$, both leading to Eq.~\ref{Eq:ABSlabDispersion}, the characteristic equation of the PhCs. 

	Then, we consider the formulae of PBC and OBC, which lead to similar spectra in Hermitian systems but have different physical meanings in essence. The former supposes all the unit-cells are connected from head to tail to preserve the energy flux along the propagation direction, enforcing $k$ to be a real number and thus $|\beta|=1$; the latter supposes the finite-size PhCs are bounded by two mirrors at which the wavefunctions are destructive interfered, which in most cases involves two Bloch states to form a standing wave. Note that the wavevectors of the two states are not required to be real numbers. To derive the relation between the two wavevectors, we write down the electric field in the FP cavity
	\begin{equation}
		E_x=Z_A m_+\beta^N_+ + Z_A m_-\beta^N_- \text{,}
		\label{Eq:ElectricField}
	\end{equation}
	and apply perfect metal boundaries to both sides~($z=0$ and $Nd$). As a result, we have
	\begin{equation}
		\beta^N_+=\beta^N_- \text{, or } k_+-k_-=\frac{2\pi M}{Nd} \text{,}
	\end{equation}
	where $M$ is an integer denoting the number of half waves in the cavity. As $N$ is sufficiently large~(thermodynamic limit), $2\pi M/(Nd)$ can be an arbitrary real number so that 
	\begin{equation}
		|\beta_+|=|\beta_-| \text{, or } Im(k_+)=Im(k_-) \text{,}
		\label{Eq:GBZ}
	\end{equation}
	meaning the exponential decay of the forward and backward waves are the same, so as to satisfy the OBCs at arbitrary length of the cavity. By solving Eq.~\ref{Eq:GBZ}, we have the GBZ of the PhCs. For previous example with Hermiticity, $k$ is always a real number for the bulk states, and therefore GBZ overlaps with the Brillouin zone (BZ). Note that Eq.~\ref{Eq:GBZ} does not work for the edge states that are self destructively interfered.

	\begin{figure}
	\centering
	\includegraphics{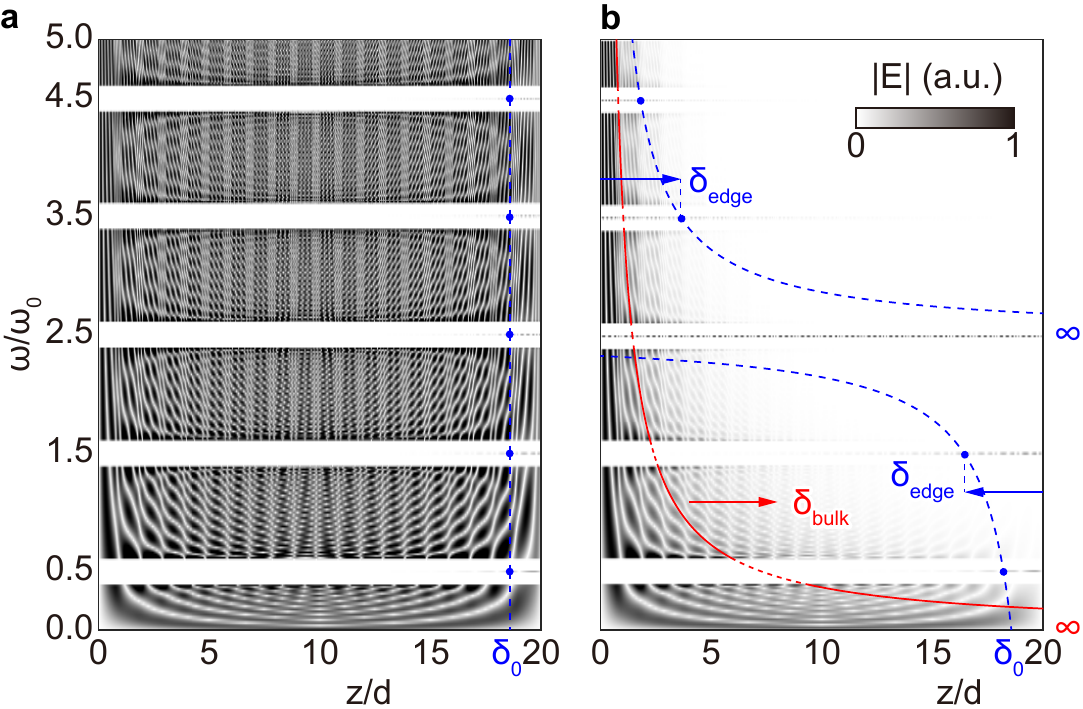}
	\caption{
		Mode profiles of eigen-states in a piece of non-Hermitian PhCs with 20 unit-cells. 
		(a) $\kappa=0$. All bulk states are extensive in the cavity, while all edge states are intrinsically localized at the right end with penetration depth $\delta_0$. 
		(b) $\kappa=1/(2.5\omega_0\delta_0d_B)$. All bulk states are localized leftwards. The edge states are localized rightwards as $0<\omega<2.5\omega_0$ and leftwards as $\omega>2.5\omega_0$. At $\omega=2.5\omega_0$, the third edge state is delocalized. The penetration depths of all states are on three branches of hyperbolic curves, described by Eqs.~\ref{Eq:BulkDelta} and \ref{Eq:EdgeDelta}, respectively.
	}
	\label{Fig:ScaleLaw}
\end{figure}

	\section{NHSE and DESs in anomalous PT photonic material}
	Based on the TMM, the NHSE can be achieved in a tricky way. Without loss of generality, we multiply both sides of Eq.~\ref{Eq:DispersionRelation} by a nonzero complex coefficient $e^{-gd}$, which is 
	\begin{equation}
		\tilde{\Psi}_B(e^{-gd}\textbf{T}_B)\tilde{\Psi}_B^{-1}
		\tilde{\Psi}_A\textbf{T}_A\tilde{\Psi}_A^{-1}
		\Psi= (e^{-gd}\beta)\Psi \text{,}
		\label{Eq:Trick}
	\end{equation}
	and suppose $\textbf{T}_{B'}=e^{-gd}\textbf{T}_B$ and $\beta'=e^{-gd}\beta$ to have a new layer B$'$ and generalized phase shift $\beta'$ to satisfy Eqs.~\ref{Eq:GBZ} and \ref{Eq:Trick} under OBC. Correspondingly, $n_{B'\pm}=n_B \pm ig/(\omega d_B)$ and $k'=k+ig$. The transform does not affect $\omega$ and $\Psi$, so the OBC spectra and the Zak phase~\cite{PhysRevLett.62.2747} are preserved; the only change is the dispersion relation between $\omega$ and $k'$. Especially, we note that as $Re(g)\neq0$, $|\beta'|\neq1$, all the Bloch states become evanescent, resulting in the NHSE. The decay rate is $Im(k')=Re(g)$ for the bulk states, and $Im(k')=Re(g)+g_0$ for the edge states. Substituting the new refractive indices into Eq.~\ref{Eq:MaterialMatrix}, we have
	\begin{equation}
		\mathcal{M}=\begin{bmatrix} \frac{n_++n_-}{2Z} & \frac{n_+-n_-}{2} \\ \frac{n_+-n_-}{2} & \frac{(n_++n_-)Z}{2}\end{bmatrix}=\begin{bmatrix} \epsilon & i\kappa \\ i\kappa & \mu \end{bmatrix}\text{,}
		\label{Eq:APTM}
	\end{equation}
	where the chirality term $\kappa=g/(\omega d_B)$. Correspondingly, the penetration depth of bulk states and edge states are 
	\begin{equation}
		\delta_\text{bulk}=\frac{1}{|d_B Re(\kappa) \omega|}
		\label{Eq:BulkDelta}
	\end{equation}
	and
	\begin{equation}
		\delta_\text{edge}=\frac{1}{|d_B Re(\kappa)\omega - 1/\delta_0|} \text{, } 
		\label{Eq:EdgeDelta}
	\end{equation}
	respectively. One can see that the penetration depth is tuned by $\kappa$ and $\omega$. For $Re(\kappa)>0$, all states tend to be localized leftwards. As the frequency increases, the localizations of bulk states become stronger in a linear pattern, leading to smaller penetration depths that are inversely proportional to frequency. For the edge states, it takes a range of frequency from $\omega=0$ to $1/(\delta_0d_BRe(\kappa))$ to offset the intrinsic rightward localizations, where the penetration depth increases from $\delta_0$ to infinity. By choosing proper values of $\kappa$, the penetration depth happens to be infinite for some edge states, giving rise to the DES. As shown in Fig.~\ref{Fig:ScaleLaw}, we plot the mode profiles of each eigen-state for $\kappa=0$ and $\kappa=1/(2.5\omega_0\delta_0d_B)$, the latter has DES at the third bandgap. For each state we mark the penetration depth by solid points where the wavefunction is $1/e$ decayed from the boundary. As shown in Fig.~\ref{Fig:ScaleLaw}(b), all the points are set on the hyperbolic curves satisfying either Eqs.~\ref{Eq:BulkDelta} or \ref{Eq:EdgeDelta}. The third edge state is exactly at the asymptotic line of two branches, doubly confirming the delocalization. As a comparison in Fig.~\ref{Fig:ScaleLaw}(a), the system is Hermitian and the bulk states are extensive in the real space, and every edge state has $\delta_0$ skin depth to the right end. 

	\begin{table}[ht]
	\centering
	\renewcommand\arraystretch{1.0}
	\begin{tabular}{c|c|c|c}
		\hline \hline
		\modtab{1mm}{1mm}{Symmetry} 
		& \modtab{1mm}{1mm}{Wavefunction} 
		& \modtab{1mm}{1mm}{Material matrix} 
		& \modtab{1mm}{1mm}{Constitutive parameters} \\ \hline
		\modtab{1mm}{0mm}{$\mathcal{H}$} 
		& $[\begin{smallmatrix} E_x \\ H_y \end{smallmatrix}]\!\!\rightarrow\!\![\begin{smallmatrix} E_x^*,H_y^* \end{smallmatrix}]$ 
		& $\mathcal{M}=\mathcal{M}^\dagger$ 
		& \modtab{1mm}{0.5mm}{
			$\epsilon=\epsilon^*$, $\mu=\mu^*$ \\ 
			$\xi=\zeta^*$ } 
		\\ \hline
		\modtab{1mm}{0mm}{$\mathcal{T}$} 
		& $[\begin{smallmatrix} E_x \\ H_y \end{smallmatrix}]\!\!\rightarrow\!\![\begin{smallmatrix} E_x^* \\ -H_y^* \end{smallmatrix}]$ 
		& $\mathcal{M}=\sigma_3\mathcal{M}^*\sigma_3$ 
		& \modtab{1mm}{0.5mm}{
			$\epsilon=\epsilon^*$, $\mu=\mu^*$ \\ 
			$\xi=-\xi^*$, $\zeta=-\zeta^*$ }
		\\ \hline
		\modtab{1mm}{0mm}{$\mathcal{P}$} 
		& $[\begin{smallmatrix} E_x \\ H_y \end{smallmatrix}]\!\!\rightarrow\!\![\begin{smallmatrix} -E_x \\ H_y \end{smallmatrix}]$ 
		& $\mathcal{M}=\sigma_3\mathcal{M}\sigma_3$ 
		& \modtab{1mm}{0.5mm}{
			$\xi=\zeta=0$ }
		\\ \hline
		\modtab{1mm}{0mm}{$\mathcal{R}=\mathcal{HT}$} 
		& $[\begin{smallmatrix} E_x \\ H_y \end{smallmatrix}]\!\!\rightarrow\!\![\begin{smallmatrix} E_x,-H_y \end{smallmatrix}]$  
		& $\mathcal{M}=\sigma_3\mathcal{M}^T\sigma_3$ 
		& \modtab{1mm}{0.5mm}{
			$\xi=-\zeta$ }
		\\ \hline
		\modtab{1mm}{0mm}{$\mathcal{PT}$} 
		& $[\begin{smallmatrix} E_x \\ H_y \end{smallmatrix}]\!\!\rightarrow\!\![\begin{smallmatrix} -E_x^* \\ -H_y^* \end{smallmatrix}]$ 
		& $\mathcal{M}=\mathcal{M}^*$ 
		& \modtab{1mm}{0.5mm}{
			$\epsilon=\epsilon^*$, $\mu=\mu^*$ \\ 
			$\xi=\xi^*$, $\zeta=\zeta^*$ } 
		\\ \hline
		\modtab{1mm}{0mm}{$\mathcal{HP}$} 
		& $[\begin{smallmatrix} E_x \\ H_y \end{smallmatrix}]\!\!\rightarrow\!\![\begin{smallmatrix} -E_x^*,H_y^* \end{smallmatrix}]$ 
		& $\mathcal{M}=\sigma_3\mathcal{M}^\dagger\sigma_3$ 
		& \modtab{1mm}{0.5mm}{
			$\epsilon=\epsilon^*$, $\mu=\mu^*$ \\ 
			$\xi=-\zeta^*$ } 
		\\ \hline
		\modtab{1mm}{0mm}{$\mathcal{HPT}$} 
		& $[\begin{smallmatrix} E_x \\ H_y \end{smallmatrix}]\!\!\rightarrow\!\![\begin{smallmatrix} -E_x^*,-H_y^* \end{smallmatrix}]$ 
		& $\mathcal{M}=\mathcal{M}^T$ 
		& \modtab{1mm}{0.5mm}{
			$\xi=\zeta$ } \\ \hline
		\hline 
	\end{tabular}
	\caption{
		Representations of Hermiticity, TRS, inversion symmetry, and some combinations of them. Note $\mathcal{R}$ is the reciprocity operator. Constraints on the material matrix and constitutive parameters are also listed. 
	}
	\label{Tab:APT}
\end{table}

	\begin{figure}
	\centering
	\includegraphics{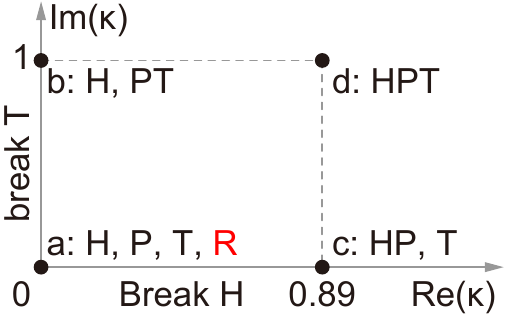}
	\caption{
		Complex plane of $\kappa$, where we pick four points~(corresponding to Fig.~\ref{Fig:FourCases}(a-d)) and label the symmetries in each case, respectively. 
	}
	\label{Fig:Kappa}
\end{figure}

	\begin{figure}
	\centering
	\includegraphics{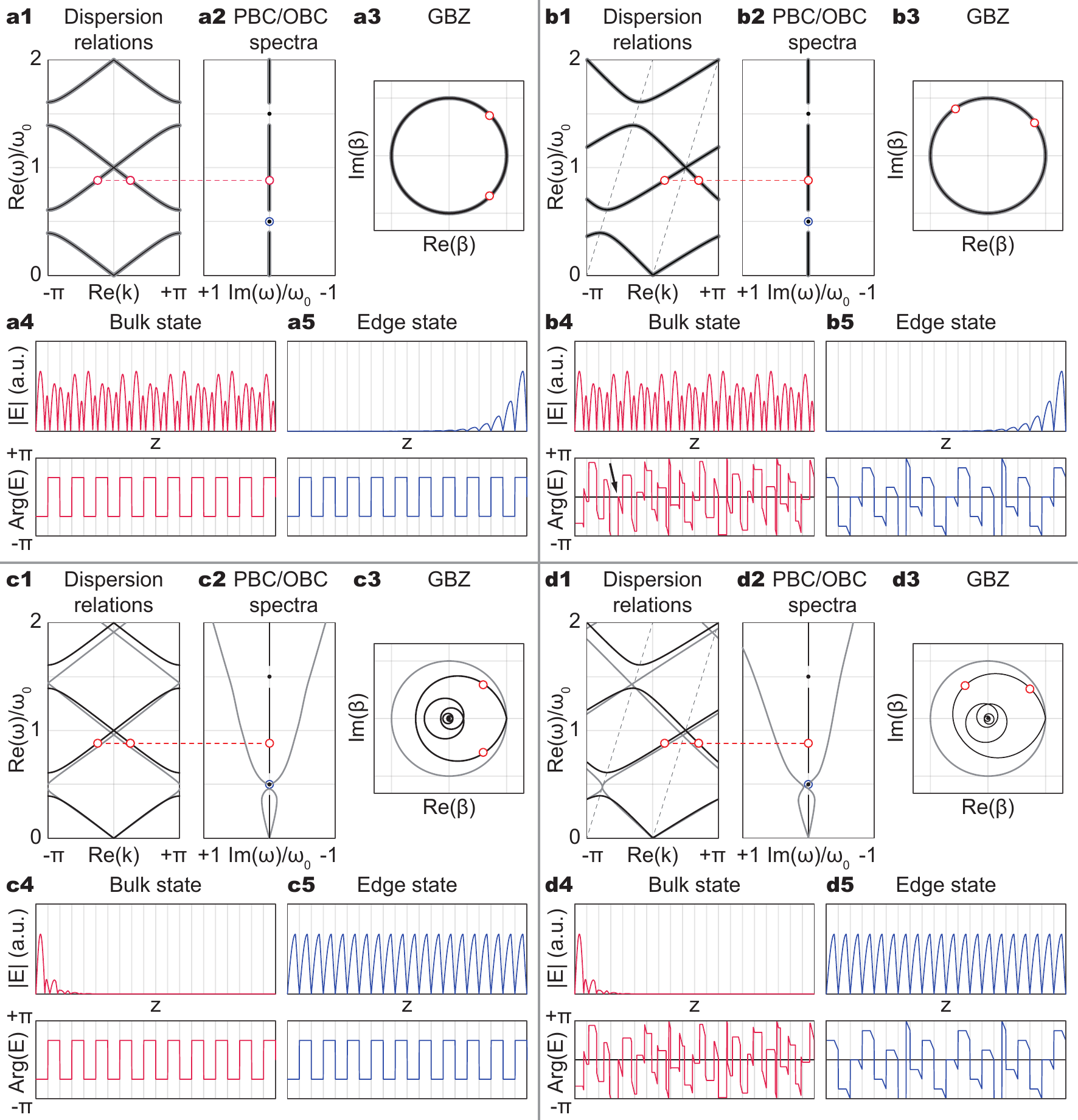}		
	\caption{
		Band diagrams of four cases of $\kappa$ in Fig.~\ref{Fig:Kappa}. 
		(a) $\kappa=0$, same as Fig.~\ref{Fig:PhC}. (a1, a2) Due to the Hermiticity, the PBC/OBC spectra are rigorous real numbers and almost the same, except discrete edge states in the bandgaps. Empty circles in (a2) denotes the standing wave under OBC that consists of two states under PBC. (a3) Complete overlap of BZ and GBZ. (a4) Mode profile of a bulk state. (a5) Mode profile of the first edge state localized at the right end. 
		(b) $\kappa=i$. The system remains Hermitian (b1) with an offset of wavevectors proportional to frequency, denoted by the dashed line. (b4, b5) Mode profiles of bulk and edge states. The phase shift is progressive along the unit-cells. 
		(c) $\kappa=-0.89$. Due to the preservation of TRS, the PBC/OBC spectra and GBZs are complex conjugate to themselves, respectively. (c1, c2) Because of the non-Hermiticity, PBC and OBC spectra no long overlap. The first edge state locates at the contour of PBC spectra, indicating the delocalization. (c3) All the GBZs lie inside the BZ, implying (c4) the leftward localization. (d5) DES. 
		(e) $\kappa=-0.89+i$. (e1, e2) Despite the breaking of TRS, the first edge state is still at the contour of PBC spectra. (e3) GBZs of each band is no longer complex conjugate to themselves, but rotates around the original point of $\beta$. (d4, d5) Mode profiles with NHSE, DES and progressive phases.  
	}
	\label{Fig:FourCases}
\end{figure}

	From Eq.~\ref{Eq:APTM} we see, the material matrix is transpose symmetric, meaning the material of layer B$'$ satisfies the anomalous parity-time~($\mathcal{HPT}$)~\cite{PhysRevLett.125.186802} symmetry. Here, $\mathcal{H}$, $\mathcal{P}$, and $\mathcal{T}$ denote the Hermiticity, the inversion symmetry, and the time-reversal symmetry (TRS) operators, respectively; the anomalous parity-time symmetry is the multiplication of the three symmetry operators. 
	
	To prove material B$'$ satisfies $\mathcal{HPT}$ symmetry, we refer to the symmetry analysis. Given a system described by the Maxwell's equations $\mathcal{D}\Psi=\omega\mathcal{M}\Psi$ and a symmetry operator $\mathcal{A}$, then the system after transform $\mathcal{A}$ is $(\mathcal{A}\mathcal{D}\mathcal{A}^{-1})(\mathcal{A}\Psi)=(\mathcal{A}\omega\mathcal{A}^{-1})(\mathcal{A}\mathcal{M}\mathcal{A}^{-1})(\mathcal{A}\Psi)$. If the system is $\mathcal{A}$ invariant, then $\mathcal{M}=\mathcal{A}\mathcal{M}\mathcal{A}^{-1}$. As shown in Column 2-4 in Tab.~\ref{Tab:APT}, the Hermiticity operator $\mathcal{H}$ performs the Hermitian conjugation to the wavefunction, turning a right eigen-state to a left eigen-state; the TRS operator is anti-unitary and flips the sign of the magnetic field; the inversion operator is unitary and flips the sign of the electric field. Restrictions on the material matrix invariant to the symmetries are listed in Row 3. With the three fundamental symmetries~(respectively commutative), we list some combinations of them to judge the symmetry satisfaction of material B$'$. As shown in the complex plane of $\kappa$ in Fig.~\ref{Fig:Kappa}, the material is $\mathcal{HPT}$ symmetric over the entire plane, and so long as $\kappa\neq0$, the material is nonreciprocal. As $\kappa$ is an imaginary number, the system is Hermitian and $\mathcal{PT}$ symmetric; as $\kappa$ is a real number, the system satisfies TRS and anomalous inversion~($\mathcal{HP}$) symmetry. 
	
	\section{Tuning penetration depth of the edge states by chirality}
	To show how the non-Hermitian skin effect and edge states are tuned by the chirality term, we choose four cases of $\kappa$~(corresponding to four points in Fig.~\ref{Fig:Kappa}) to calculate the dispersion relations, the PBC/OBC spectra, the mode profiles of bulk and edge states, and the GBZs of the first four bands from zero frequency, respectively, as shown in Fig.~\ref{Fig:FourCases}. We start with $\kappa=0$, the origin of the $\kappa$ space, as shown in Fig.~\ref{Fig:FourCases}(a), which is already mentioned in Fig.~\ref{Fig:PhC}. From Fig.~\ref{Fig:FourCases}(a) to Fig.~\ref{Fig:FourCases}(b), where $\kappa$ turns into an imaginary number, the wavevector is attached with a bias $(Im(\kappa) d_B/d) \omega$, performing an Arnold transform to the dispersion relation, as shown in Fig.~\ref{Fig:FourCases}(b1). Unlike the $\pm\pi/2$ phase in Fig.~\ref{Fig:FourCases}(a4, a5), the phases in Fig.~\ref{Fig:FourCases}(b4, b5) are progressively shifted in a linear pattern, since the Bloch states comprising the standing wave do not belong to a pair of opposite wavevectors, as shown in Fig.~\ref{Fig:FourCases}(b1). Due to the Hermiticity of PhCs, the PBC/OBC spectra is preserved, and the GBZ overlaps with BZ, which is the unit-circle in the complex plane of $\beta$.
	
	Next, we introduce NHSE by allowing $Re(\kappa)=0.89$, which delocalize the first edge state at $\omega=\omega_0/2$. As shown in Fig.~\ref{Fig:FourCases}(c), where $\kappa$ is purely a real number, both the forward and backward wavevectors are attached with an identical imaginary part to localize the states to the left boundary. Due to the non-Hermiticity, PBC and OBC spectra split apart, while locates exactly at the contour of PBC spectra, meaning this state does not need to add a nonzero imaginary part to the wavevector to satisfy the OBC, implying it is a DES, as shown in Fig.~\ref{Fig:FourCases}(c5). We note that the material has $\mathcal{HP}$ symmetry~\cite{PhysRevB.103.205205}, which is reported in another work on DES realized by tight-binding model~\cite{PhysRevB.103.195414}. Besides, due to the TRS, the PBC/OBC spectra and the GBZ of each band overlap with the complex conjugate of themselves, respectively.
	
	Finally, as $\kappa$ is generally a complex number, such as $\kappa=0.89+i$ in Fig.~\ref{Fig:FourCases}(d), both real and imaginary part of the wavevector is tuned that result in both the progressive phase and the NHSE. Despite the breaking of TRS, the edge state is still on the contour of PBC spectra and therefore is delocalized.  

	\section{Discussions}
	We have thus proposed to realize the NHSE and the DES in realistic photonic crystals with $\mathcal{HPT}$ symmetry, based on the Maxwell’s equations. We reveal that, in general, the penetration depth of the eigen-state is inversely proportional to the chirality and frequency. With a careful choice of the chirality, the offset between the penetration depth and the intrinsic decay length of the edge state leads to the fully delocalized in-gap edge state that can be useful for laser applications. Besides, chiral material with $\mathcal{HPT}$ symmetry guarantees the non-reciprocity, which provides an alternative way to break reciprocity~\cite{PhysRevLett.124.257403}. 
	
	\paragraph{Funding.}---
	The work at Zhejiang University was sponsored by the National Natural Science Foundation of China (NNSFC) under Grants No. 61625502, No. 62175215, No. 11961141010, No. 61975176, and No. U19A2054, the Top-Notch Young Talents Program of China, and the Fundamental Research Funds for the Central Universities. 
	\paragraph{Disclosures.}---
	The authors declare no conflicts of interest.
	\paragraph{Data Availability.}---
	Data underlying the results presented in this paper are not publicly available at this time but may be obtained from the authors upon reasonable request.

	\nocite{apsrev41Control}
	\bibliographystyle{apsrev4}
	\bibliography{titleon,DES}
	
\end{document}